\documentclass[a4,aps,prl,twocolumn,groupeadress,lengthcheck]{revtex4}

\usepackage{amsmath}

\usepackage{graphicx}
\usepackage{bm}
\usepackage{latexsym}
\usepackage{array}
\usepackage{color}
\usepackage[dvipsnames]{xcolor}
\usepackage[normalem]{ulem}
\usepackage{hyperref}
\newcolumntype{P}[1]{>{\centering\arraybackslash}p{#1}}
\newcolumntype{M}[1]{>{\centering\arraybackslash}m{#1}}
\input{epsf}

\begin{document}
\title{Puzzling Bubble Rise Speed Increase in Dense Granular Suspensions}
\author{Christopher Madec$^1$, Briva\"el Collin$^1$, J. John Soundar Jerome$^2$, Sylvain Joubaud$^{1,3}$}
\affiliation{$^1$ Univ Lyon, ENS de Lyon, Univ Claude Bernard, CNRS, Laboratoire de Physique, $46$ all\'{e}e d'Italie, $69364$, Lyon, France}
\affiliation{$^2$ Universit\'{e} de Lyon, Universit\'{e} Claude Bernard Lyon$1$, Laboratoire de M\'{e}canique des Fluides et d'Acoustique, CNRS, UMR $5509$, Boulevard $11$ Novembre, $69622$ Villeurbanne CEDEX, Lyon, France}
\affiliation{$^3$ Institut Universitaire de France (IUF), $1$ rue Descartes, $75005$, Paris, France}
\date{\today}

\begin{abstract}
We present an anomalous experimental observation on the rising speed of air bubbles in a Hele-Shaw cell containing a suspension of spherical, neutrally buoyant, non-Brownian particles. Strikingly, bubbles rise faster in suspensions as compared to particle-less liquids of the same effective viscosity. By carefully measuring this bubble speed increase at various particle volume fraction and via velocity field imaging, we demonstrate that this strange bubble dynamics is linked to a reduction in the bulk dissipation rate. A good match between our experimental data and computations based on Suspension Balance Model illustrates that the underlying mechanism for this dissipation-rate-deficit is related to a nonuniform particle distribution in the direction perpendicular to the channel walls due to shear-induced particle migration.

\end{abstract}

\maketitle

{\em Introduction --} From their spontaneous birth, then throughout their wobbly life and even during their violent death, bubbles~\citep{prosperetti2004bubbles} strongly influence mass and heat transfer in many modern engineering techniques and industrial processes such as mixing in chemical reactors with bubble columns and cooling systems, aerosol transfer, contaminant removal in alloy melts, flows in petroleum industry, carbon sequestration, ship hydrodynamics, to name just a few~\citep{clift2005bubbles}. They are also of huge importance in biological and geophysical phenomena. Indeed, they have justly occupied a large body of modern research in fluid mechanics~\citep{wegener1973spherical, fabre1992modeling, feng1997nonlinear, magnaudet2000motion, bodner2012bubble, tripathi2015dynamics, Dollet2018}. Historically, investigations on bubble dynamics were primarily focused on the two-phases, namely, bubble and the surrounding liquid, to understand sound generation, terminal velocity, wake, shape and path instabilities of a single bubble at relatively large Reynolds numbers in an unbounded media \citep{levich1962physicochemical, haberman1954experimental, saffman1956rise, ellingsen2001rise, mougin2001path, veldhuis2008shape}, and/or a Hele-Shaw cell \citep{TaylorSaffman1959, Maxworthy1986, Tanveer1986, kopf1988bubble, kelley1997path,filella2015oscillatory,roig2012dynamics}.

Nevertheless, solid particles are often present alongside bubbles in most of the aformentioned applications. In this context, along with advances in suspensions rheology \citep{GuazzelliPouliquen2018} and increasing interests on the role of particles on liquid films \citep{Gans_SoftMatter2019, Sauret_PRF2019, PalmaLhuissier_JFM2019} and bubbles in microchannels~\citep{YuStone2018}, studies on single bubble dynamics in a suspension constitute a simple yet important class of research to understand free boundary problems in multiphase fluid dynamics. Surprisingly, only a few works~\citep{luo1997rise, liang2013bubble, HooshyarPRL2013, YuStone2018} have considered this basic problem. While a recent work~\citep{HooshyarPRL2013} suggests that, in the absence of walls, particles greatly smaller than bubbles do not influence bubble dynamics, there are still many open questions: What is the effect of the bubble {and the particle size} on the particle-laden liquid, and vice versa, under strong confinement? So, what is the bubble speed? Also, how do bubbles interact with each other in such flows? In this Letter, we try to answer the former two questions by reporting on a novel result from a model experiment to investigate the rising motion of a single, isolated bubble through a neutrally buoyant liquid-particle suspension in a Hele-Shaw cell.

\begin{figure*}
\includegraphics[width=.99\textwidth]{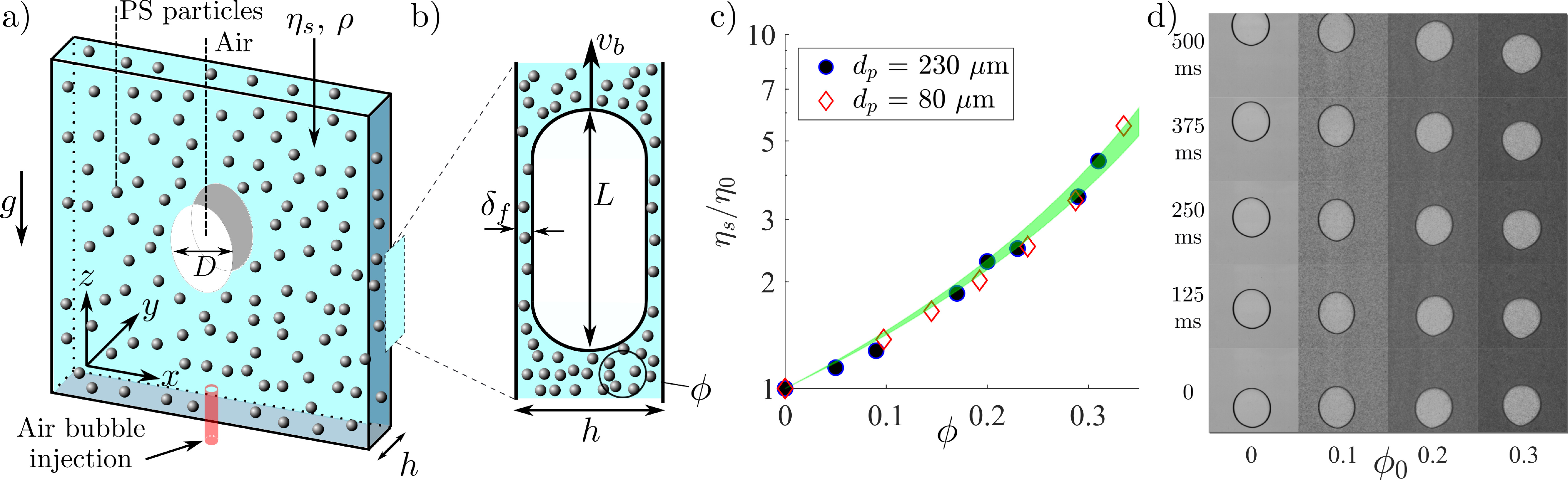}
\caption{\label{fig:Experimental_device} Rise of an air bubble in a Hele-Shaw cell filled with a viscous suspension composed of a  water and UCON$\scriptsize{\texttrademark}$ mixture and polystyrene (PS) particles. Schematic of the (a) front view and (b) side view of the setup. (c) Normalized suspension viscosity $\hat{\eta}(\phi) \equiv \eta_s(\phi_0)/\eta_0$. Symbols represent experimental data and the green region is Eq.~(\ref{eq:viscosity}) with $\phi_c \in [0.55-0.62]$. (d) Time-lapse imaging of rising air bubbles ($d_{b}=12.5$ $\pm$ 0.1 mm) in a suspension ($d_p = 230$~$\mu$m) for different bulk packing fractions $\phi_0$.}
\end{figure*}
{\em Experimental setup --} It consists of a Hele-Shaw cell made of two $20$ $\times$ $25$ cm glass plates which are separated by a small gap of $h =2.30 \pm 0.05$ mm~[Figs.~\ref{fig:Experimental_device}(a)-\ref{fig:Experimental_device}(b)]. The cell is gently filled from the top with a non-Brownian suspension composed of spherical, quasimonodispersed polystyrene (PS) particles of mean diameter $d_p$ ($230$~$\pm$~$10$~$\mu$m or $80$~$\pm$~$3$~$\mu$m) and a viscous Newtonian liquid (UCON$\scriptsize{\texttrademark}$ and water mixture of controllable dynamic viscosity $\eta_0$). Special care was taken when each suspension sample was prepared by properly mixing PS particles ($\rho = 1057 \pm 2$ kg m$^{-3}$) with a suspending liquid of same density in order to obtain a homogeneous mix of particle volume fraction $\phi_0$ (the ratio of the volume of particles to the total volume). In particular, the suspension was always left on a roller-mixer setup to avoid particles from settling or floating in case of a weak density mismatch, if any. The suspension bulk viscosity $\eta_s{(\phi_0)}$ is systematically measured using a rheometer (Malvern Kinexus) at  shear rates of $0.1$--$10$ $s^{-1}$ and at $25$ $^\circ$C. Two common rotational geometries, namely, a Taylor-Couette set-up and a parallel-plate rheometer with different gaps were used to assure repeatability and validity of all viscosity measurements, avoiding any bias from shear-induced migration~\cite{gadala1980shear}. As seen in Fig.~\ref{fig:Experimental_device}(c), a good agreement is obtained with the Maron-Pierce formula~\citep{MaronPierce1956, GuazzelliPouliquen2018} (see the Supplemental Material~\cite{Supplemental}):
\begin{equation}\label{eq:viscosity}
\hat{\eta}(\phi) \equiv \frac{\eta_s ({\phi})}{\eta_0} = \left(1-\dfrac{\phi}{\phi_c}\right)^{-2},
\end{equation}
if the random close packing concentration $\phi_c$ ranges between $0.55$ and $0.625$ following commonly used values in the literature~\cite{GuazzelliPouliquen2018}. Note that the cell gap to particle {diameter} ratio is about {10}, or higher, and so confinement effects on suspension rheology are {expected to be} small {for $\phi_0<0.3$}~\citep{Davit_EPL2008,YeoMaxey_PRL2010, Peyla_EPL2011, Gallier_JFM2016,Doyeux}.

\begin{figure}[t]
\includegraphics[width=.49\textwidth]{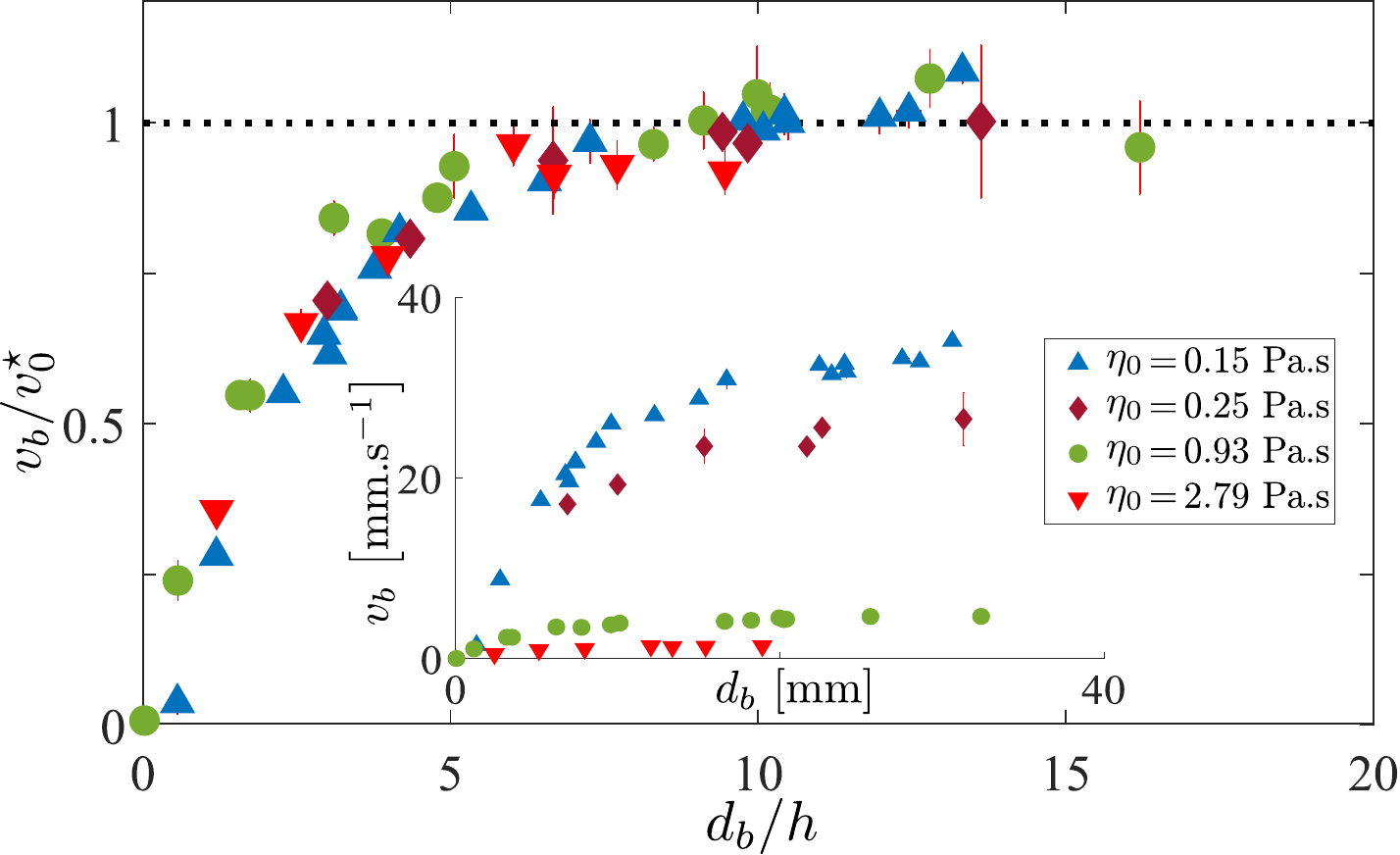}
\caption{\label{fig:Viscosity}Raw experimental datasets for different particle-less water and UCON mixtures (inset) collapse onto a unique master curve when plotted as normalized bubble rise velocity $v_b/v_0^{\star}$ vs the normalized diameter $d_b/h$. The $y$ axis error bars are computed through the fluctuations of $v_b$ in time.}
\end{figure}

A single bubble is released from a capillary tube at the bottom of the cell. The bubble motion is recorded using a high-resolution CCD camera producing $2048 \times 2048$ pixel images at $20$ to $80$ fps. As sketched in Fig.~\ref{fig:Experimental_device}(b), bubbles extend over almost the full thickness of the Hele-Shaw cell while a thin liquid layer of an average thickness $\delta_f$ is observed. For particle-free liquids, we measured $\delta_f=200-400$~$\mu$m using a profilometer as in the classical Bretherton's law~\citep{BrethertonJFM1961}.  Bubbles are slightly elongated and are characterized by two geometrical parameters, namely, the equivalent diameter $d_b = \sqrt{4{\mathcal A_b}/\pi}$ with ${\mathcal A}_b$ the measured area of the bubbles and the aspect ratio $L/D$~[Figs.~\ref{fig:Experimental_device}(a)-\ref{fig:Experimental_device}(b)].  In all our experiments, bubbles rise at a steady velocity, denoted $v_b$, just after a few millimeters above the release point.

{\em Bubble rise in Newtonian liquids --} At first, we consider the evolution of $v_b$ as a function of the bubble diameter $d_b$ for four different water and UCON mixtures with distinct viscosity $\eta_0$ in the absence of particles. As expected, at a given $d_b$, the bubble rise is slower in a much viscous liquid (see inset, Fig.~\ref{fig:Viscosity}). At a fixed $\eta_0$, the bubble rising speed increases as the bubble size increases while inset Fig.~\ref{fig:Viscosity} further suggests that the bubble speed tends towards a constant upper bound when $d_b/h > 6$. This is in accordance with classical theoretical works~\citep{TaylorSaffman1959, Tanveer1986, Maxworthy1986}, which predict a maximum speed limit for flat bubbles, say $v_0^{\star}$, in Hele-Shaw cell containing Newtonian liquids given by
\begin{equation}\label{eq:Maxworthy}
v_0^{\star} \equiv v^{\star}(\eta_{0}) = \dfrac{\Delta \rho g h^2}{12 \eta_{0}} \left( \dfrac{L}{D} \right),
\end{equation}
where $\Delta \rho$ is the density difference between the surrounding liquid (here, $\rho_0$) and the gas (air) in the bubble. As clearly demonstrated by Fig.~\ref{fig:Viscosity}, the Saffman-Taylor-Maxworthy velocity $v_0^{\star}$ is, within experimental error bars and without any adjustable parameters, the right velocity scale for all data in the case of Newtonian liquids without particles. Indeed, for all bubbles here, $v_b$ is inversely proportional to the liquid viscosity $\eta_0$.

\begin{figure*}
\includegraphics[width=.99\textwidth]{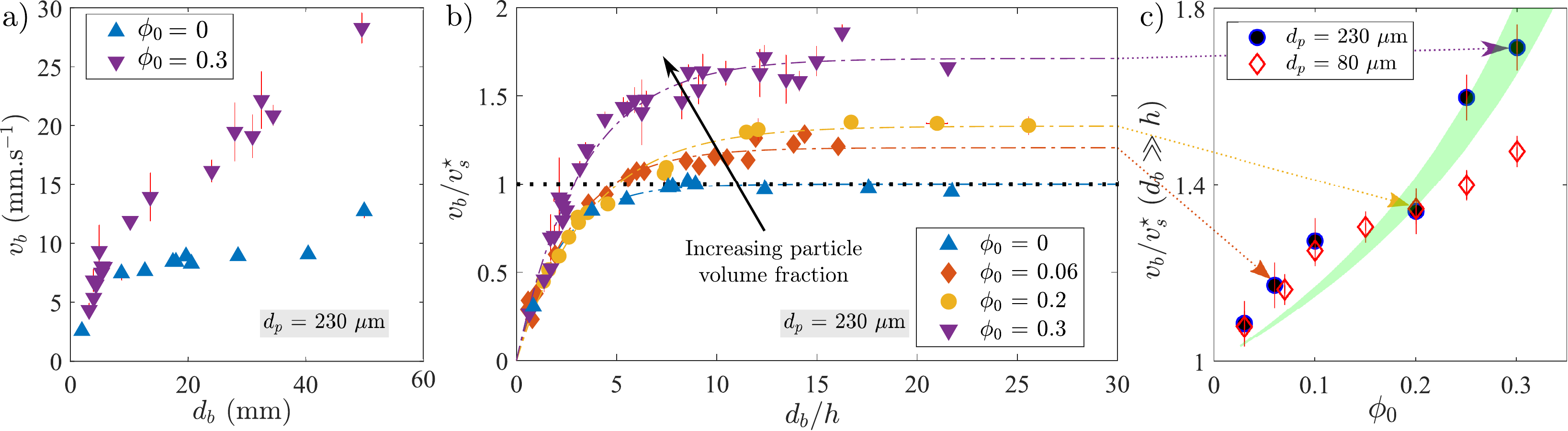}
\caption{\label{fig:chrono_velocity_230}a) Comparison of bubble rise velocity $v_b$ between a liquid with particles and a pure liquid of same bulk viscosity, $0.53$~Pa.s showing that bubble rises faster in the former. Note that a plateau can be observed if the velocity is divided by the aspect ratio $L/D$ [Eq.~(\ref{eq:Maxworthy})]. b) Normalized bubble velocity~$v_b/v_s^\star$ vs normalized diameter $d_b/h$ illustrates the anomalous bubble rise speed increase at various $\phi_0$. Empirical fits (dashed-dotted lines) are used to calculate $v_b/v_s^\star$ when $d_b \gg h$, whose value is then plotted in c) as a function of suspension volume fraction $\phi_0$ for both $d_p = 230$ and $80$~$\mu$m. Symbols represent experimental data and the green region Eq.~(\ref{eq:alpha_factor_rise}) using a volume fraction profile given by Eq.~(\ref{eq:profile_MaronPierceSBM}) with $\phi_c \in [0.55-0.62]$.}
\end{figure*}

{\em Bubble rise in suspensions --} Figure~\ref{fig:Experimental_device}(d) presents chronophotographs of bubble motion in suspensions of same UCON and water mixtures but for different bulk packing fraction $\phi_0$. For all cases, the packing fraction is uniform in the $XZ$ plane around the bubble, while inside the bubble, particles are clearly visible in the gap between the bubble and channel walls, as can be expected since $d_p \lesssim \delta_f$~\citep{YuStone2018,Gans_SoftMatter2019, Sauret_PRF2019, PalmaLhuissier_JFM2019}. At a fixed diameter $d_b = 12.5$~mm and $\eta_0=0.18$~Pa.s, a bubble rises slower in a denser suspension since the bulk viscosity $\eta_s(\phi_0)$ increases [Fig.~\ref{fig:Experimental_device}(c)] with $\phi_0$. On the other hand, Fig.~\ref{fig:chrono_velocity_230}(a) emphasizes that a bubble in a non-Brownian suspension of same bulk viscosity as a particle-free liquid rises puzzlingly faster. To further investigate this anomalous behavior in neutrally  buoyant suspension, we measure the bubble rise velocity by systematically varying the bulk suspension volume fraction $\phi_0$ between $0\%-30$\%. We then take, as before, the corresponding Saffman-Taylor-Maxworthy velocity $v_s^{\star} \equiv v^{\star}\left[\eta_s(\phi_0)\right]$ [Eq.~(\ref{eq:Maxworthy})], and plot in Fig.~\ref{fig:chrono_velocity_230}(b) the renormalized bubble velocity $v_b/v_s^{\star}$ against the nondimensional bubble diameter $d_b/h$. Similar to the case of a Newtonian liquid without particles [Fig.~\ref{fig:Viscosity}], it increases steadily with the bubble size and reaches a limiting value when $d_b \gg h$. On the contrary, all data for the renormalized bubble speed in suspensions ($\phi_0 > 0$) do not collapse and the maximum bubble speed in the limit ${d_b \gg h}$ increases with particle concentration and gets as high as $\sim1.8 v_s^\star$ at $\phi_0 = 30$\% [see Fig.~\ref{fig:chrono_velocity_230}(c)]. This not only confirms that bubble speed augmentation is observed even when $\phi_0$ is small and but also indicates that, for two particle-laden liquids of same bulk viscosity, the bubble speed increases with bulk volume fraction $\phi_0$.

Figure~\ref{fig:chrono_velocity_230}(c) also displays data from experiments with  smaller PS particles ($d_p=80$~$\mu$m). No {dependency} with $d_p$ is observed for $\phi_0 \leq 20$\%, beyond which the velocity increase is larger for larger particle mean diameter.

\begin{figure}
\includegraphics[width=.49\textwidth]{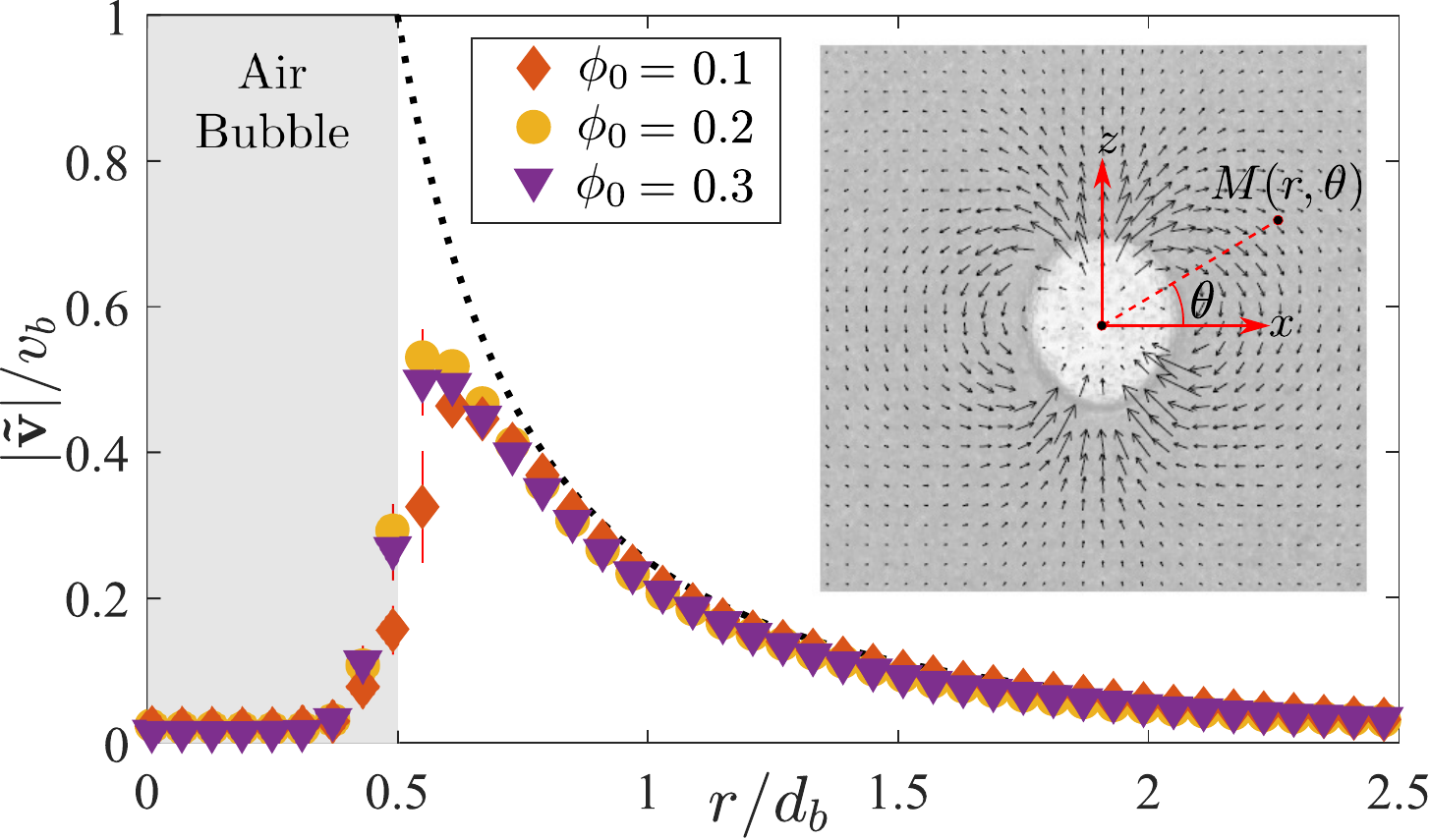}
\caption{\label{fig:PIV}Magnitude of the local velocity around a bubble of size $d_b = 25$ mm for various packing fraction $\phi_0$. Dotted line represents the theoretical dipolar field, $({v_b}{d_b}/{4}){r}^{-2}$. The $y$ axis {error bars} are computed from standard error over different temporal realizations of $\vert\tilde{\mathbf{v}}/v_b\vert$. Inset: instantaneous velocity field from PIV measurements ($\phi_0 = 0.10$, $d_p=230$~$\mu$m).}
\end{figure}

For further insight, the local suspension hydrodynamics around the bubble is investigated by exploiting the motion of the suspension texture as captured by the camera (see Movies in the Supplemental Material~\cite{Supplemental}). The local thickness-averaged velocity field of the suspension, denoted by $\tilde{\mathbf{v}}(r,\theta)$, is then computed using a classical cross-correlation method, implemented in the open-source software UVMAT with windows of typical size 13 $\times$ 13 pixels at a 15\% overlap. Thereby, Fig.~\ref{fig:PIV} shows that the magnitude of the velocity field $\vert\mathbf{\tilde{v}}\vert$ decreases with the distance $r$ from the bubble center following a dipolar field $({v_b}{d_b^2}/{4}){r}^{-2}$. Although, such a potential flow field is typical for Newtonian liquids in Hele-Shaw cell~\citep{hele1898flow}, to the authors' knowledge, this is the first experimental evidence of an incompressible, potential flow around a bubble in a Hele-Shaw cell containing dense granular suspensions. Consequently, no matter $\phi_0$, the resulting dissipation should be predominantly due to friction losses across the cell thickness. While the packing fraction does not depend on $x$ and $z$ [see the snapshots, Fig.~\ref{fig:Experimental_device}(d) and Movies in the Supplemental Material~\cite{Supplemental}], it could be nonuniform over the thickness of the tank. It is precisely in this context that we revisit the energetic arguments of Maxworthy~\citep{Maxworthy1986} and extend it to account for a possible nonuniform particle distribution across the Hele-Shaw cell.

The bubble speed results from a balance between $\mathcal{P}_B$, the injected power due to buoyancy to rise a bubble and the viscous dissipation rate. For an elliptical bubble rising at a speed $v_b$ in the limit $d_b \gg h$, the power due to buoyancy $\mathcal{P}_B = \Delta \rho g \left( \pi D L h/4\right) v_b$ is the same with or without particles. And for the dissipation rate, it is convenient to distinguish several contributions: (i) $\mathcal{P}_{\delta}$ due to friction between the bubble and the channel wall and (ii) $\mathcal{P}_{\ell}$ arising from the bulk suspension motion set in by the bubble. As demonstrated in~\citep{Keiser2018} for a pure liquid, since the viscosity of the gas inside the bubble is negligible when compared with liquid viscosity, the liquid velocity almost vanishes over the entire width $\delta_f$ so that the lubrication film does not play a significant role. In the presence of particles, the depth-averaged velocity in the gap between the bubble and the wall is close to zero as inferred from Fig.~\ref{fig:PIV} in the region $r<d_b/2$. Therefore, lubrication should be negligibly small compared to the viscous dissipation rate $\mathcal{P}_{\ell}$ due to suspension motion across the channel. If $d_b\gg h$,
\begin{equation}
\mathcal{P}_{\ell}=\displaystyle\iiint_{\mathrm{cell}} \ \eta_s(y) \left\vert \frac{\partial \mathbf{v}}{\partial y}\right\vert^2 \mathrm{d}\tau,
\end{equation}
where the suspension velocity field $\mathbf{v}(r,\theta,y)$ should obey the incompressible viscous suspension flow $\bm{\nabla}P = \frac{\partial}{\partial y}\left(\eta_s(y)\frac{\partial \mathbf{v}}{\partial y}\right)$ since the bubble Reynolds number $\mathrm{Re}_b~\ll~1$ (about $0.1$ or less in all our experiments)~\citep{GuazzelliPouliquen2018}. Note that the local suspension viscosity $\eta_s(y)$ is taken to vary only in the $y$ direction like the local packing fraction $\phi(y)$. By taking a parallel flow assumption across the cell, we note that the pressure field $P$ depends only on $(r, \theta)$ while the thickness-averaged velocity $\tilde{\mathbf{v}}(r, \theta) = \frac{1}{h}\int_{-h/2}^{+h/2}\mathbf{v}(r,\theta, y)\mathrm{d}y$ is then simply given by the experimentally illustrated dipolar field. After some algebra (see the Supplemental Material), the bulk dissipation rate reads $\mathcal{P}_\ell  =  {3\pi D^2 \eta_s(\phi_0) v_b^{2}}/{\alpha(\phi_0) h}$ with
\begin{eqnarray}
\alpha(\phi_0) & = & \frac{3}{2} \displaystyle\int_{-1}^{1}\hat{y}^2\frac{\hat{\eta}(\phi_0)}{\hat{\eta}\left[\phi(\hat{y})\right]}\mathrm{d}\hat{y},
\label{eq:alpha_factor_rise}
\end{eqnarray}
when $d_b \gg h$ (here, $\hat{y} = 2y/h$).  At equilibrium, as for a Newtonian liquid~\citep{Maxworthy1986}, $\mathcal{P}_B = \mathcal{P}_\ell$. Thereby, we see that $\alpha(\phi_0)$ is precisely the velocity overshoot, $v_b/v_s^\star \ (d_b \gg h)$. Also, the factor $\alpha(\phi_0)$ is equal to unity for a particle-less Newtonian liquid $\left[\phi(\hat{y})=\phi_0= 0\right]$ or a uniform profile $\left[\phi(\hat{y})=\phi_0\neq 0\right]$.

In granular suspensions, particle-particle collisions contribute to nonzero normal stress differences and lead to the well-known shear-induced particle migration~\citep{leighton1987shear, GuazzelliPouliquen2018}. The so-called Suspension Balance Model (SBM)~\cite{nott1994pressure} provides a reasonably good estimate of the local particle fraction evolution $\phi(\mathbf{x},t)$ in many experiments~\citep{koh1994experimental,LyonJFM1998,snook2016dynamics, sarabian2019} and simulations \citep{morris1999curvilinear, nott1994pressure, yeo2011numerical}. For a fully developed channel flow, SBM with Maron-Pierce viscosity [Eq.~(\ref{eq:viscosity})] gives
\begin{eqnarray}
\phi(\hat{y}) = 
	\begin{cases} 
		\phi_c & 0 \leq \vert \hat{y} \vert  \leq \beta, \\
		\phi_c\sqrt{\beta/\hat{y}} & \beta \leq \vert \hat{y} \vert  \leq 1,
   	\end{cases}
	\label{eq:profile_MaronPierceSBM}
\end{eqnarray}
where $\sqrt{\beta} = 1 -\sqrt{ 1 - \phi_0/\phi_c}$.
\begin{figure}
\includegraphics[width=.49\textwidth] {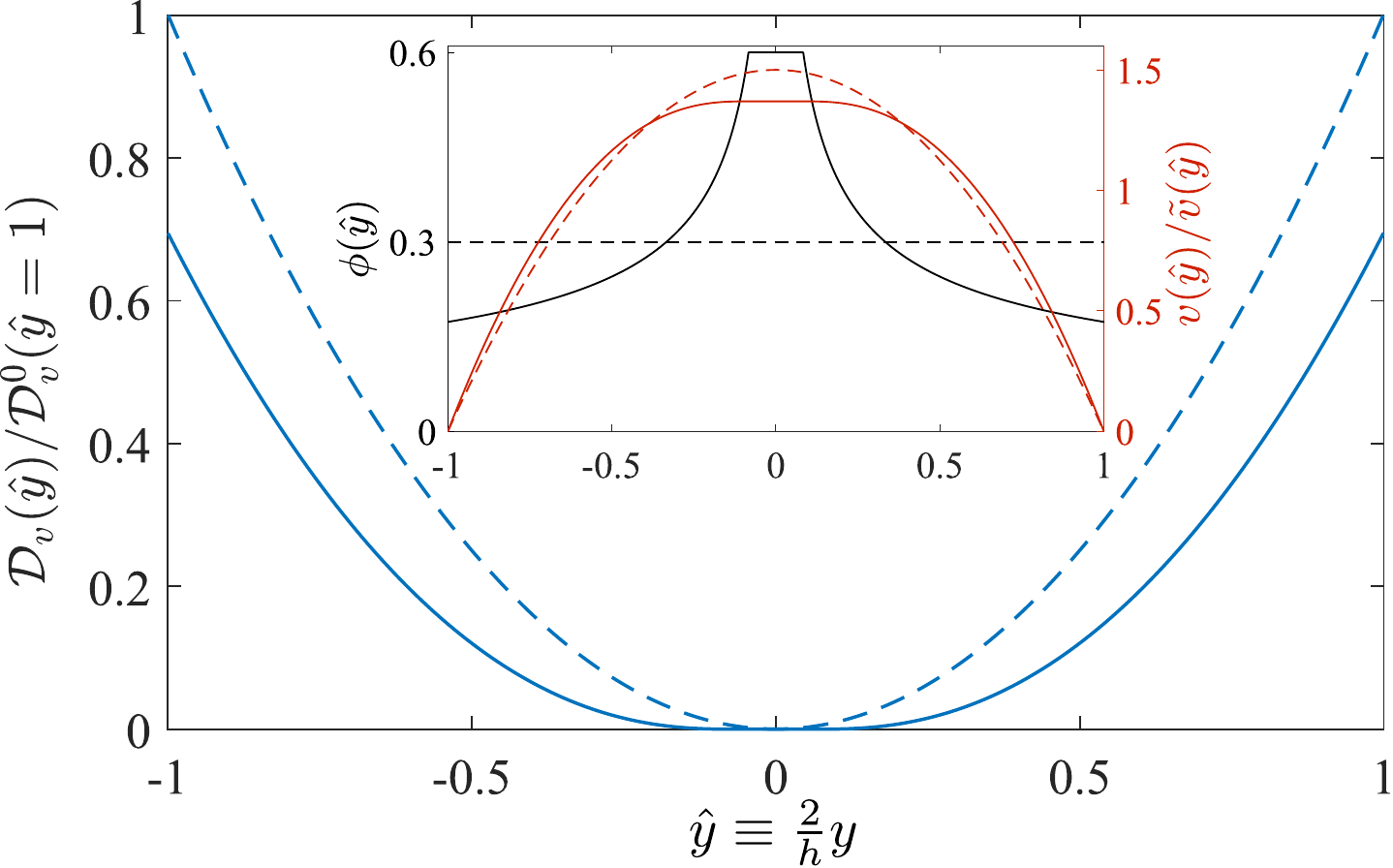}
\caption{\label{fig:Vinf}Local dissipation rate ${\mathcal D}_v(\hat{y})=\eta_s(\hat{y})\left\vert {\partial \mathbf{v}}/{\partial \hat{y}}\right\vert^2$ across the channel for a suspension with (continuous line) and without (dashed line) particle migration. Here, ${\mathcal D}_v^0(\hat{y} = 1)$ is the dissipation rate at the wall in the case of uniform particle distribution $\phi(\hat{y}) = \phi_0$. Clearly, ${\mathcal D}_v(\hat{y})$ is smaller in the case with shear-induced migration. Inset: corresponding velocity (red) and volume fraction (black) profiles.}
\end{figure}
For example, this volume fraction profile and its corresponding channel velocity along with the local dissipation rate per unit volume ${\mathcal D}_v(\hat{y})=\eta_s(\hat{y})\left\vert {\partial \mathbf{v}}/{\partial \hat{y}}\right\vert^2$ are provided in Fig.~\ref{fig:Vinf}, when $\phi_0 = 30$\%.  By comparison with the classical channel flow profiles when $\phi(\hat{y}) = \phi_0$ (dashed lines) of the same bulk flow rate, it is evident from Fig.~\ref{fig:Vinf} that the local dissipation rate across the channel is reduced by the presence of this nonuniform particle distribution. Indeed, the dissipation is localized in the strong velocity gradient region and thus mainly in the region where $\phi(\hat{y}) < \phi_0$ and thus, leading to a reduced total dissipation rate. We now proceed to a direct comparison between the SBM-based $\alpha(\phi_0)$ and the experimentally measured velocity overshoot $v_b/v_s^\star \ (d_b \gg h)$ in Fig.~\ref{fig:chrono_velocity_230}(c) without any adjustable parameters. Indeed, the results from our simple model match very well with all experimental data presented here. This result confirms that dissipation-deficit via particle migration is the principal mechanism underlying the bubble speed augmentation when the bulk particle concentration is increased. We point out, however, that the steady-state SBM solution does not capture the experimentally observed dependence of $\alpha(\phi_0)$ on $d_p$ in Fig.~\ref{fig:chrono_velocity_230}(c)  when $\phi_0 > 20$\%. This could be due to time-dependent migration dynamics, the exact particle distribution across the channel, the difference in rheology due to lateral confinement, the interaction between the particles and the lubrication film which are beyond the scope of the present Letter.

\vspace{11pt}

{\em Conclusion --} Unlike the Newtonian case, bubble rise speed in suspensions contained in a Hele-Shaw cell cannot simply be inversely proportional to the suspension viscosity $\eta_s(\phi_0)$ alone, even when the bulk volume fraction is as small as $3$\%. Indeed, our measurements clearly show that bubbles rise faster in suspensions than in a pure liquid of the same bulk viscosity. We elucidate the key ingredient for this anomalous bubble speed increase to be the particle migration which, when properly incorporated via the well-known Suspension Balance Model~\cite{nott1994pressure}, accounts for the reduced bulk dissipation rate of the suspension flow around the rising bubble. 

These results open multiple paths for further investigations to understand bubbles in dispersed media. Firstly, our findings point out that bubbles and particles strongly interact when bubbles are confined whereas previous results on unbounded systems~\cite{HooshyarPRL2013} suggest that bubbles are not influenced by the particles. So, we hope that our work motivates research to elucidate the transition between these two regimes. Moreover, our work hints that there could be a rich phenomenology for the case with larger bubble Reynolds number, for example, by revisiting the scaling $v_b\propto\sqrt{gd_b}$~\cite{roig2012dynamics}, shape and path instabilities, to name a few, in the presence of suspension.  While, in perspective, numerical modeling to capture multiscale interactions between solid particles, suspending liquid, and fluid-fluid interfaces is difficult due to complex rheology of suspensions, it is also a challenging endeavor to experimentally identify the physics of interactions between multiple bubbles.
As suggested by the anomalous bubble rise speed in our Hele-Shaw setup Fig.~[\ref{fig:chrono_velocity_230}(c)], both simulation and experiments to quantitatively estimate the local packing fraction and the 3D-particle dynamics in the cell would be crucial. In addition, experiments to explore confinement effects on bubble and particle interactions might be of paramount interest for future investigations.

We acknowledge M. Moulin for experimental help. We thank S. Dagois-Bohy, E. Guazzelli, P. Jouin, S. Manneville, J.-P. Matas and V. Vidal for useful discussions. This work was supported by the LABEX iMUST (Grant No. ANR-10-LABX-0064) of Universit\'{e} de Lyon, within the program ``Investissements d'Avenir'' (Grant No. ANR-11-IDEX-0007) operated by the French National Research Agency (ANR). It has been achieved thanks to the resources of PSMN from ENS de Lyon.

\bibliography{biblioPRL_BubbleSpeedHeleShaw}

\providecommand{\noopsort}[1]{}\providecommand{\singleletter}[1]{#1}%
\begin{thebibliography}{50}
\expandafter\ifx\csname natexlab\endcsname\relax\def\natexlab#1{#1}\fi
\expandafter\ifx\csname bibnamefont\endcsname\relax
  \def\bibnamefont#1{#1}\fi
\expandafter\ifx\csname bibfnamefont\endcsname\relax
  \def\bibfnamefont#1{#1}\fi
\expandafter\ifx\csname citenamefont\endcsname\relax
  \def\citenamefont#1{#1}\fi
\expandafter\ifx\csname url\endcsname\relax
  \def\url#1{\texttt{#1}}\fi
\expandafter\ifx\csname urlprefix\endcsname\relax\def\urlprefix{URL }\fi
\providecommand{\bibinfo}[2]{#2}
\providecommand{\eprint}[2][]{\url{#2}}

\bibitem[{\citenamefont{Prosperetti}(2004)}]{prosperetti2004bubbles}
\bibinfo{author}{\bibfnamefont{A.}~\bibnamefont{Prosperetti}},
  \bibinfo{journal}{\href{https://doi.org/10.1063/1.1695308}{Phys. Fluids}}
  \textbf{\bibinfo{volume}{16}}, \bibinfo{pages}{1852} (\bibinfo{year}{2004}).

\bibitem[{\citenamefont{Clift et~al.}(2005)\citenamefont{Clift, Grace, and
  Weber}}]{clift2005bubbles}
\bibinfo{author}{\bibfnamefont{R.}~\bibnamefont{Clift}},
  \bibinfo{author}{\bibfnamefont{J.~R.} \bibnamefont{Grace}}, \bibnamefont{and}
  \bibinfo{author}{\bibfnamefont{M.~E.} \bibnamefont{Weber}},
  \emph{\bibinfo{title}{Bubbles, drops, and particles}}
  (\bibinfo{publisher}{Courier Corporation}, \bibinfo{year}{2005}).

\bibitem[{\citenamefont{Wegener and Parlange}(1973)}]{wegener1973spherical}
\bibinfo{author}{\bibfnamefont{P.~P.} \bibnamefont{Wegener}} \bibnamefont{and}
  \bibinfo{author}{\bibfnamefont{J.-Y.} \bibnamefont{Parlange}},
  \bibinfo{journal}{\href{https://doi.org/10.1146/annurev.fl.05.010173.000455}{Ann.
  Rev. of Fluid Mech.}} \textbf{\bibinfo{volume}{5}}, \bibinfo{pages}{79}
  (\bibinfo{year}{1973}).

\bibitem[{\citenamefont{Fabre and Lin{\'e}}(1992)}]{fabre1992modeling}
\bibinfo{author}{\bibfnamefont{J.}~\bibnamefont{Fabre}} \bibnamefont{and}
  \bibinfo{author}{\bibfnamefont{A.}~\bibnamefont{Lin{\'e}}},
  \bibinfo{journal}{\href{https://doi.org/10.1146/annurev.fl.24.010192.000321}{Ann.
  Rev. of Fluid Mech.}} \textbf{\bibinfo{volume}{24}}, \bibinfo{pages}{21}
  (\bibinfo{year}{1992}).

\bibitem[{\citenamefont{Feng and Leal}(1997)}]{feng1997nonlinear}
\bibinfo{author}{\bibfnamefont{Z.}~\bibnamefont{Feng}} \bibnamefont{and}
  \bibinfo{author}{\bibfnamefont{L.}~\bibnamefont{Leal}},
  \bibinfo{journal}{\href{https://doi.org/10.1146/annurev.fluid.29.1.201}{Ann.
  Rev. of Fluid Mech.}} \textbf{\bibinfo{volume}{29}}, \bibinfo{pages}{201}
  (\bibinfo{year}{1997}).

\bibitem[{\citenamefont{Magnaudet and Eames}(2000)}]{magnaudet2000motion}
\bibinfo{author}{\bibfnamefont{J.}~\bibnamefont{Magnaudet}} \bibnamefont{and}
  \bibinfo{author}{\bibfnamefont{I.}~\bibnamefont{Eames}},
  \bibinfo{journal}{\href{https://doi.org/10.1146/annurev.fluid.32.1.659}{Ann.
  Rev. of Fluid Mech.}} \textbf{\bibinfo{volume}{32}}, \bibinfo{pages}{659}
  (\bibinfo{year}{2000}).

\bibitem[{\citenamefont{Prosperetti}(1992)}]{bodner2012bubble}
\bibinfo{author}{\bibfnamefont{A.}~\bibnamefont{Prosperetti}}, in
  \emph{\bibinfo{booktitle}{Theoretical and Applied Mechanics}}, edited by
  \bibinfo{editor}{\bibfnamefont{S.}~\bibnamefont{Bodner}},
  \bibinfo{editor}{\bibfnamefont{S.}~\bibnamefont{J.}},
  \bibinfo{editor}{\bibfnamefont{S.}~\bibnamefont{A.}}, \bibnamefont{and}
  \bibinfo{editor}{\bibfnamefont{H.}~\bibnamefont{Z.}}
  (\bibinfo{publisher}{Elsevier, Amsterdam}, \bibinfo{year}{1992}), pp.
  \bibinfo{pages}{355--369}.

\bibitem[{\citenamefont{Tripathi et~al.}(2015)\citenamefont{Tripathi, Sahu, and
  Govindarajan}}]{tripathi2015dynamics}
\bibinfo{author}{\bibfnamefont{M.~K.} \bibnamefont{Tripathi}},
  \bibinfo{author}{\bibfnamefont{K.~C.} \bibnamefont{Sahu}}, \bibnamefont{and}
  \bibinfo{author}{\bibfnamefont{R.}~\bibnamefont{Govindarajan}},
  \bibinfo{journal}{\href{https://doi.org/10.1038/ncomms7268}{Nat. Commun.}}
  \textbf{\bibinfo{volume}{6}}, \bibinfo{pages}{6268} (\bibinfo{year}{2015}).

\bibitem[{\citenamefont{Dollet et~al.}(2019)\citenamefont{Dollet, Marmottant,
  and Garbin}}]{Dollet2018}
\bibinfo{author}{\bibfnamefont{B.}~\bibnamefont{Dollet}},
  \bibinfo{author}{\bibfnamefont{P.}~\bibnamefont{Marmottant}},
  \bibnamefont{and} \bibinfo{author}{\bibfnamefont{V.}~\bibnamefont{Garbin}},
  \bibinfo{journal}{\href{https://doi.org/10.1146/annurev-fluid-010518-040352}{Ann.
  Rev. of Fluid Mech.}} \textbf{\bibinfo{volume}{51}}, \bibinfo{pages}{331}
  (\bibinfo{year}{2019}).

\bibitem[{\citenamefont{Levich}(1962)}]{levich1962physicochemical}
\bibinfo{author}{\bibfnamefont{V.~G.} \bibnamefont{Levich}},
  \emph{\bibinfo{title}{Physicochemical hydrodynamics}}
  (\bibinfo{publisher}{Prentice-Hall Inc.}, \bibinfo{year}{1962}).

\bibitem[{\citenamefont{Haberman}(1954)}]{haberman1954experimental}
\bibinfo{author}{\bibfnamefont{W.~L.} \bibnamefont{Haberman}},
  \bibinfo{journal}{Trans. ASCE} \textbf{\bibinfo{volume}{2799}},
  \bibinfo{pages}{227} (\bibinfo{year}{1954}).

\bibitem[{\citenamefont{Saffman}(1956)}]{saffman1956rise}
\bibinfo{author}{\bibfnamefont{P.}~\bibnamefont{Saffman}},
  \bibinfo{journal}{\href{https://doi.org/10.1017/S0022112056000159}{J. Fluid
  Mech.}} \textbf{\bibinfo{volume}{1}}, \bibinfo{pages}{249}
  (\bibinfo{year}{1956}).

\bibitem[{\citenamefont{Ellingsen and Risso}(2001)}]{ellingsen2001rise}
\bibinfo{author}{\bibfnamefont{K.}~\bibnamefont{Ellingsen}} \bibnamefont{and}
  \bibinfo{author}{\bibfnamefont{F.}~\bibnamefont{Risso}},
  \bibinfo{journal}{\href{https://doi.org/10.1017/S0022112001004761}{J. Fluid
  Mech.}} \textbf{\bibinfo{volume}{440}}, \bibinfo{pages}{235}
  (\bibinfo{year}{2001}).

\bibitem[{\citenamefont{Mougin and Magnaudet}(2001)}]{mougin2001path}
\bibinfo{author}{\bibfnamefont{G.}~\bibnamefont{Mougin}} \bibnamefont{and}
  \bibinfo{author}{\bibfnamefont{J.}~\bibnamefont{Magnaudet}},
  \bibinfo{journal}{\href{https://doi.org/10.1103/PhysRevLett.88.014502}{Phys.
  Rev. Lett.}} \textbf{\bibinfo{volume}{88}}, \bibinfo{pages}{014502}
  (\bibinfo{year}{2001}).

\bibitem[{\citenamefont{Veldhuis et~al.}(2008)\citenamefont{Veldhuis,
  Biesheuvel, and Van~Wijngaarden}}]{veldhuis2008shape}
\bibinfo{author}{\bibfnamefont{C.}~\bibnamefont{Veldhuis}},
  \bibinfo{author}{\bibfnamefont{A.}~\bibnamefont{Biesheuvel}},
  \bibnamefont{and}
  \bibinfo{author}{\bibfnamefont{L.}~\bibnamefont{Van~Wijngaarden}},
  \bibinfo{journal}{\href{https://doi.org/10.1063/1.2911042}{Phys. Fluids}}
  \textbf{\bibinfo{volume}{20}}, \bibinfo{pages}{040705}
  (\bibinfo{year}{2008}).

\bibitem[{\citenamefont{Taylor and Saffman}(1959)}]{TaylorSaffman1959}
\bibinfo{author}{\bibfnamefont{G.}~\bibnamefont{Taylor}} \bibnamefont{and}
  \bibinfo{author}{\bibfnamefont{P.}~\bibnamefont{Saffman}},
  \bibinfo{journal}{\href{https://doi.org/10.1093/qjmam/12.3.265}{Q. J. Mech.
  Appl. Math.}} \textbf{\bibinfo{volume}{12}}, \bibinfo{pages}{265}
  (\bibinfo{year}{1959}).

\bibitem[{\citenamefont{Maxworthy}(1986)}]{Maxworthy1986}
\bibinfo{author}{\bibfnamefont{T.}~\bibnamefont{Maxworthy}},
  \bibinfo{journal}{\href{https://doi.org/10.1017/S002211208600109X}{J. Fluid
  Mech.}} \textbf{\bibinfo{volume}{173}}, \bibinfo{pages}{95}
  (\bibinfo{year}{1986}).

\bibitem[{\citenamefont{Tanveer}(1986)}]{Tanveer1986}
\bibinfo{author}{\bibfnamefont{S.}~\bibnamefont{Tanveer}},
  \bibinfo{journal}{\href{https://doi.org/10.1063/1.865831}{Phys. Fluids}}
  \textbf{\bibinfo{volume}{29}}, \bibinfo{pages}{3537} (\bibinfo{year}{1986}).

\bibitem[{\citenamefont{Kopf-Sill and Homsy}(1988)}]{kopf1988bubble}
\bibinfo{author}{\bibfnamefont{A.~R.} \bibnamefont{Kopf-Sill}}
  \bibnamefont{and} \bibinfo{author}{\bibfnamefont{G.}~\bibnamefont{Homsy}},
  \bibinfo{journal}{\href{https://doi.org/10.1063/1.866566}{Phys. Fluids}}
  \textbf{\bibinfo{volume}{31}}, \bibinfo{pages}{18} (\bibinfo{year}{1988}).

\bibitem[{\citenamefont{Kelley and Wu}(1997)}]{kelley1997path}
\bibinfo{author}{\bibfnamefont{E.}~\bibnamefont{Kelley}} \bibnamefont{and}
  \bibinfo{author}{\bibfnamefont{M.}~\bibnamefont{Wu}},
  \bibinfo{journal}{\href{https://doi.org/10.1103/PhysRevLett.79.1265}{Phys.
  Rev. Lett.}} \textbf{\bibinfo{volume}{79}}, \bibinfo{pages}{1265}
  (\bibinfo{year}{1997}).

\bibitem[{\citenamefont{Filella et~al.}(2015)\citenamefont{Filella, Ern, and
  Roig}}]{filella2015oscillatory}
\bibinfo{author}{\bibfnamefont{A.}~\bibnamefont{Filella}},
  \bibinfo{author}{\bibfnamefont{P.}~\bibnamefont{Ern}}, \bibnamefont{and}
  \bibinfo{author}{\bibfnamefont{V.}~\bibnamefont{Roig}},
  \bibinfo{journal}{\href{https://doi.org/10.1017/jfm.2015.355}{J. Fluid
  Mech.}} \textbf{\bibinfo{volume}{778}}, \bibinfo{pages}{60}
  (\bibinfo{year}{2015}).

\bibitem[{\citenamefont{Roig et~al.}(2012)\citenamefont{Roig, Roudet, Risso,
  and Billet}}]{roig2012dynamics}
\bibinfo{author}{\bibfnamefont{V.}~\bibnamefont{Roig}},
  \bibinfo{author}{\bibfnamefont{M.}~\bibnamefont{Roudet}},
  \bibinfo{author}{\bibfnamefont{F.}~\bibnamefont{Risso}}, \bibnamefont{and}
  \bibinfo{author}{\bibfnamefont{A.-M.} \bibnamefont{Billet}},
  \bibinfo{journal}{\href{https://doi.org/10.1017/jfm.2012.289}{J. Fluid
  Mech.}} \textbf{\bibinfo{volume}{707}}, \bibinfo{pages}{444}
  (\bibinfo{year}{2012}).

\bibitem[{\citenamefont{Guazzelli and
  Pouliquen}(2018)}]{GuazzelliPouliquen2018}
\bibinfo{author}{\bibfnamefont{E.}~\bibnamefont{Guazzelli}} \bibnamefont{and}
  \bibinfo{author}{\bibfnamefont{O.}~\bibnamefont{Pouliquen}},
  \bibinfo{journal}{\href{https://doi.org/10.1017/jfm.2018.548}{J. Fluid
  Mech.}} \textbf{\bibinfo{volume}{852}}, \bibinfo{pages}{P1}
  (\bibinfo{year}{2018}).

\bibitem[{\citenamefont{Gans et~al.}(2019)\citenamefont{Gans, Dressaire,
  Colnet, Saingier, Bazant, and Sauret}}]{Gans_SoftMatter2019}
\bibinfo{author}{\bibfnamefont{A.}~\bibnamefont{Gans}},
  \bibinfo{author}{\bibfnamefont{E.}~\bibnamefont{Dressaire}},
  \bibinfo{author}{\bibfnamefont{B.}~\bibnamefont{Colnet}},
  \bibinfo{author}{\bibfnamefont{G.}~\bibnamefont{Saingier}},
  \bibinfo{author}{\bibfnamefont{M.~Z.} \bibnamefont{Bazant}},
  \bibnamefont{and} \bibinfo{author}{\bibfnamefont{A.}~\bibnamefont{Sauret}},
  \bibinfo{journal}{\href{https://doi.org/10.1039/C8SM01785A}{Soft matter}}
  \textbf{\bibinfo{volume}{15}}, \bibinfo{pages}{252} (\bibinfo{year}{2019}).

\bibitem[{\citenamefont{Sauret et~al.}(2019)\citenamefont{Sauret, Gans, Colnet,
  Saingier, Bazant, and Dressaire}}]{Sauret_PRF2019}
\bibinfo{author}{\bibfnamefont{A.}~\bibnamefont{Sauret}},
  \bibinfo{author}{\bibfnamefont{A.}~\bibnamefont{Gans}},
  \bibinfo{author}{\bibfnamefont{B.}~\bibnamefont{Colnet}},
  \bibinfo{author}{\bibfnamefont{G.}~\bibnamefont{Saingier}},
  \bibinfo{author}{\bibfnamefont{M.~Z.} \bibnamefont{Bazant}},
  \bibnamefont{and}
  \bibinfo{author}{\bibfnamefont{E.}~\bibnamefont{Dressaire}},
  \bibinfo{journal}{\href{https://doi.org/10.1103/PhysRevFluids.4.054303}{Physical
  Review Fluids}} \textbf{\bibinfo{volume}{4}}, \bibinfo{pages}{054303}
  (\bibinfo{year}{2019}).

\bibitem[{\citenamefont{Palma and Lhuissier}(2019)}]{PalmaLhuissier_JFM2019}
\bibinfo{author}{\bibfnamefont{S.}~\bibnamefont{Palma}} \bibnamefont{and}
  \bibinfo{author}{\bibfnamefont{H.}~\bibnamefont{Lhuissier}},
  \bibinfo{journal}{\href{https://doi.org/10.1017/jfm.2019.267}{Journal of
  Fluid Mechanics}} \textbf{\bibinfo{volume}{869}} (\bibinfo{year}{2019}).

\bibitem[{\citenamefont{Yu et~al.}(2018)\citenamefont{Yu, Khodaparast, and
  Stone}}]{YuStone2018}
\bibinfo{author}{\bibfnamefont{Y.~E.} \bibnamefont{Yu}},
  \bibinfo{author}{\bibfnamefont{S.}~\bibnamefont{Khodaparast}},
  \bibnamefont{and} \bibinfo{author}{\bibfnamefont{H.~A.} \bibnamefont{Stone}},
  \bibinfo{journal}{\href{https://doi.org/10.1063/1.5023341}{Applied Physics
  Letters}} \textbf{\bibinfo{volume}{112}}, \bibinfo{pages}{181604}
  (\bibinfo{year}{2018}).

\bibitem[{\citenamefont{Luo et~al.}(1997)\citenamefont{Luo, Zhang, Tsuchiya,
  and Fan}}]{luo1997rise}
\bibinfo{author}{\bibfnamefont{X.}~\bibnamefont{Luo}},
  \bibinfo{author}{\bibfnamefont{J.}~\bibnamefont{Zhang}},
  \bibinfo{author}{\bibfnamefont{K.}~\bibnamefont{Tsuchiya}}, \bibnamefont{and}
  \bibinfo{author}{\bibfnamefont{L.-S.} \bibnamefont{Fan}},
  \bibinfo{journal}{\href{https://doi.org/10.1016/S0009-2509(97)00215-7}{Chem.
  Eng. Sci.}} \textbf{\bibinfo{volume}{52}}, \bibinfo{pages}{3693}
  (\bibinfo{year}{1997}).

\bibitem[{\citenamefont{Liang-Shih and Tsuchiya}(2013)}]{liang2013bubble}
\bibinfo{author}{\bibfnamefont{F.}~\bibnamefont{Liang-Shih}} \bibnamefont{and}
  \bibinfo{author}{\bibfnamefont{K.}~\bibnamefont{Tsuchiya}},
  \emph{\bibinfo{title}{Bubble wake dynamics in liquids and liquid-solid
  suspensions}} (\bibinfo{publisher}{Butterworth-Heinemann},
  \bibinfo{year}{2013}).

\bibitem[{\citenamefont{Hooshyar et~al.}(2013)\citenamefont{Hooshyar, van
  Ommen, Hamersma, Sundaresan, and Mudde}}]{HooshyarPRL2013}
\bibinfo{author}{\bibfnamefont{N.}~\bibnamefont{Hooshyar}},
  \bibinfo{author}{\bibfnamefont{J.~R.} \bibnamefont{van Ommen}},
  \bibinfo{author}{\bibfnamefont{P.~J.} \bibnamefont{Hamersma}},
  \bibinfo{author}{\bibfnamefont{S.}~\bibnamefont{Sundaresan}},
  \bibnamefont{and} \bibinfo{author}{\bibfnamefont{R.~F.} \bibnamefont{Mudde}},
  \bibinfo{journal}{\href{https://doi.org/10.1103/PhysRevLett.110.244501}{Phys.
  Rev. Lett.}} \textbf{\bibinfo{volume}{110}}, \bibinfo{pages}{244501}
  (\bibinfo{year}{2013}).

\bibitem[{\citenamefont{Gadala-Maria and Acrivos}(1980)}]{gadala1980shear}
\bibinfo{author}{\bibfnamefont{F.}~\bibnamefont{Gadala-Maria}}
  \bibnamefont{and} \bibinfo{author}{\bibfnamefont{A.}~\bibnamefont{Acrivos}},
  \bibinfo{journal}{\href{https://doi.org/10.1122/1.549584}{J. Rheol}}
  \textbf{\bibinfo{volume}{24}}, \bibinfo{pages}{799} (\bibinfo{year}{1980}).

\bibitem[{\citenamefont{Maron and Pierce}(1956)}]{MaronPierce1956}
\bibinfo{author}{\bibfnamefont{S.~H.} \bibnamefont{Maron}} \bibnamefont{and}
  \bibinfo{author}{\bibfnamefont{P.~E.} \bibnamefont{Pierce}},
  \bibinfo{journal}{\href{https://doi.org/10.1016/0095-8522(56)90023-X}{J.
  Colloid Sci.}} \textbf{\bibinfo{volume}{11}}, \bibinfo{pages}{80}
  (\bibinfo{year}{1956}).

\bibitem[{\citenamefont{Davit and Peyla}(2008)}]{Davit_EPL2008}
\bibinfo{author}{\bibfnamefont{Y.}~\bibnamefont{Davit}} \bibnamefont{and}
  \bibinfo{author}{\bibfnamefont{P.}~\bibnamefont{Peyla}},
  \bibinfo{journal}{\href{https://doi.org/10.1209/0295-5075/83/64001}{Europhysics
  Letters}} \textbf{\bibinfo{volume}{83}}, \bibinfo{pages}{64001}
  (\bibinfo{year}{2008}).

\bibitem[{\citenamefont{Yeo and Maxey}(2010)}]{YeoMaxey_PRL2010}
\bibinfo{author}{\bibfnamefont{K.}~\bibnamefont{Yeo}} \bibnamefont{and}
  \bibinfo{author}{\bibfnamefont{M.~R.} \bibnamefont{Maxey}},
  \bibinfo{journal}{\href{https://link.aps.org/doi/10.1103/PhysRevE.81.051502}{Phys.
  Rev. E}} \textbf{\bibinfo{volume}{81}}, \bibinfo{pages}{051502}
  (\bibinfo{year}{2010}).

\bibitem[{\citenamefont{Peyla and Verdier}(2011)}]{Peyla_EPL2011}
\bibinfo{author}{\bibfnamefont{P.}~\bibnamefont{Peyla}} \bibnamefont{and}
  \bibinfo{author}{\bibfnamefont{C.}~\bibnamefont{Verdier}},
  \bibinfo{journal}{\href{https://doi.org/10.1209/0295-5075/94/44001}{Europhysics
  Letters}} \textbf{\bibinfo{volume}{94}}, \bibinfo{pages}{44001}
  (\bibinfo{year}{2011}).

\bibitem[{\citenamefont{Gallier et~al.}(2016)\citenamefont{Gallier, Lemaire,
  Lobry, and Peters}}]{Gallier_JFM2016}
\bibinfo{author}{\bibfnamefont{S.}~\bibnamefont{Gallier}},
  \bibinfo{author}{\bibfnamefont{E.}~\bibnamefont{Lemaire}},
  \bibinfo{author}{\bibfnamefont{L.}~\bibnamefont{Lobry}}, \bibnamefont{and}
  \bibinfo{author}{\bibfnamefont{F.}~\bibnamefont{Peters}},
  \bibinfo{journal}{\href{https://doi.org/10.1017/jfm.2016.368}{Journal of
  Fluid Mechanics}} \textbf{\bibinfo{volume}{799}}, \bibinfo{pages}{100–127}
  (\bibinfo{year}{2016}).

\bibitem[{\citenamefont{{Doyeux, Vincent and Priem, Stephane and Jibuti, Levan
  and Farutin, Alexander and Ismail, Mourad and Peyla,
  Philippe}}(2016)}]{Doyeux}
\bibinfo{author}{\bibnamefont{{Doyeux, Vincent and Priem, Stephane and Jibuti,
  Levan and Farutin, Alexander and Ismail, Mourad and Peyla, Philippe}}},
  \bibinfo{journal}{{Phys. Rev. Fluids}} \textbf{\bibinfo{volume}{1}},
  \bibinfo{pages}{043301} (\bibinfo{year}{2016}).

\bibitem[{\citenamefont{Bretherton}(1961)}]{BrethertonJFM1961}
\bibinfo{author}{\bibfnamefont{F.}~\bibnamefont{Bretherton}},
  \bibinfo{journal}{\href{https://doi.org/10.1017/S0022112061000160}{J. Fluid.
  Mech.}} \textbf{\bibinfo{volume}{10}}, \bibinfo{pages}{166}
  (\bibinfo{year}{1961}).

\bibitem[{\citenamefont{Hele-Shaw}(1898)}]{hele1898flow}
\bibinfo{author}{\bibfnamefont{H.~S.} \bibnamefont{Hele-Shaw}},
  \bibinfo{journal}{\href{https://doi.org/10.1038/058034a0}{Nature}}
  \textbf{\bibinfo{volume}{58}} (\bibinfo{year}{1898}).

\bibitem[{\citenamefont{Keiser et~al.}(2018)\citenamefont{Keiser, Jaafar, Bico,
  and Reyssat}}]{Keiser2018}
\bibinfo{author}{\bibfnamefont{L.}~\bibnamefont{Keiser}},
  \bibinfo{author}{\bibfnamefont{K.}~\bibnamefont{Jaafar}},
  \bibinfo{author}{\bibfnamefont{J.}~\bibnamefont{Bico}}, \bibnamefont{and}
  \bibinfo{author}{\bibfnamefont{E.}~\bibnamefont{Reyssat}},
  \bibinfo{journal}{\href{https://doi.org/10.1017/jfm.2018.240}{J. Fluid
  Mech.}} \textbf{\bibinfo{volume}{845}}, \bibinfo{pages}{245}
  (\bibinfo{year}{2018}).

\bibitem[{\citenamefont{Leighton and Acrivos}(1987)}]{leighton1987shear}
\bibinfo{author}{\bibfnamefont{D.}~\bibnamefont{Leighton}} \bibnamefont{and}
  \bibinfo{author}{\bibfnamefont{A.}~\bibnamefont{Acrivos}},
  \bibinfo{journal}{\href{https://doi.org/10.1017/S0022112087002155}{J. Fluid
  Mech.}} \textbf{\bibinfo{volume}{181}}, \bibinfo{pages}{415}
  (\bibinfo{year}{1987}).

\bibitem[{\citenamefont{Nott and Brady}(1994)}]{nott1994pressure}
\bibinfo{author}{\bibfnamefont{P.~R.} \bibnamefont{Nott}} \bibnamefont{and}
  \bibinfo{author}{\bibfnamefont{J.~F.} \bibnamefont{Brady}},
  \bibinfo{journal}{\href{https://doi.org/10.1017/S0022112094002326}{J. Fluid
  Mech.}} \textbf{\bibinfo{volume}{275}}, \bibinfo{pages}{157}
  (\bibinfo{year}{1994}).

\bibitem[{\citenamefont{Koh et~al.}(1994)\citenamefont{Koh, Hookham, and
  Leal}}]{koh1994experimental}
\bibinfo{author}{\bibfnamefont{C.~J.} \bibnamefont{Koh}},
  \bibinfo{author}{\bibfnamefont{P.}~\bibnamefont{Hookham}}, \bibnamefont{and}
  \bibinfo{author}{\bibfnamefont{L.~G.} \bibnamefont{Leal}},
  \bibinfo{journal}{\href{https://doi.org/10.1017/S0022112094000911}{J. Fluid
  Mech.}} \textbf{\bibinfo{volume}{266}}, \bibinfo{pages}{1}
  (\bibinfo{year}{1994}).

\bibitem[{\citenamefont{Lyon and Leal}(1998)}]{LyonJFM1998}
\bibinfo{author}{\bibfnamefont{M.~K.} \bibnamefont{Lyon}} \bibnamefont{and}
  \bibinfo{author}{\bibfnamefont{L.~G.} \bibnamefont{Leal}},
  \bibinfo{journal}{\href{https://doi.org/10.1017/S0022112098008817}{J. Fluid.
  Mech.}} \textbf{\bibinfo{volume}{363}} (\bibinfo{year}{1998}).

\bibitem[{\citenamefont{Snook et~al.}(2016)\citenamefont{Snook, Butler, and
  Guazzelli}}]{snook2016dynamics}
\bibinfo{author}{\bibfnamefont{B.}~\bibnamefont{Snook}},
  \bibinfo{author}{\bibfnamefont{J.~E.} \bibnamefont{Butler}},
  \bibnamefont{and}
  \bibinfo{author}{\bibfnamefont{{\'E}.}~\bibnamefont{Guazzelli}},
  \bibinfo{journal}{\href{https://doi.org/10.1017/jfm.2015.645}{J. Fluid
  Mech.}} \textbf{\bibinfo{volume}{786}}, \bibinfo{pages}{128}
  (\bibinfo{year}{2016}).

\bibitem[{\citenamefont{Sarabian et~al.}(2019)\citenamefont{Sarabian,
  Firouznia, Metzger, and Hormozi}}]{sarabian2019}
\bibinfo{author}{\bibfnamefont{M.}~\bibnamefont{Sarabian}},
  \bibinfo{author}{\bibfnamefont{M.}~\bibnamefont{Firouznia}},
  \bibinfo{author}{\bibfnamefont{B.}~\bibnamefont{Metzger}}, \bibnamefont{and}
  \bibinfo{author}{\bibfnamefont{S.}~\bibnamefont{Hormozi}},
  \bibinfo{journal}{\href{https://doi.org/10.1017/jfm.2018.982}{J. Fluid.
  Mech.}} \textbf{\bibinfo{volume}{862}}, \bibinfo{pages}{659–671}
  (\bibinfo{year}{2019}).

\bibitem[{\citenamefont{Morris and Boulay}(1999)}]{morris1999curvilinear}
\bibinfo{author}{\bibfnamefont{J.~F.} \bibnamefont{Morris}} \bibnamefont{and}
  \bibinfo{author}{\bibfnamefont{F.}~\bibnamefont{Boulay}},
  \bibinfo{journal}{\href{https://doi.org/10.1122/1.551021}{J. Rheol}}
  \textbf{\bibinfo{volume}{43}}, \bibinfo{pages}{1213} (\bibinfo{year}{1999}).

\bibitem[{\citenamefont{Yeo and Maxey}(2011)}]{yeo2011numerical}
\bibinfo{author}{\bibfnamefont{K.}~\bibnamefont{Yeo}} \bibnamefont{and}
  \bibinfo{author}{\bibfnamefont{M.~R.} \bibnamefont{Maxey}},
  \bibinfo{journal}{\href{https://doi.org/10.1017/jfm.2011.241}{J. Fluid
  Mech.}} \textbf{\bibinfo{volume}{682}}, \bibinfo{pages}{491}
  (\bibinfo{year}{2011}).

\bibitem[{\citenamefont{Boyer et~al.}(2011)\citenamefont{Boyer, Guazzelli, and
  Pouliquen}}]{BoyerPRL2011}
\bibinfo{author}{\bibfnamefont{F.}~\bibnamefont{Boyer}},
  \bibinfo{author}{\bibfnamefont{{\'E}.}~\bibnamefont{Guazzelli}},
  \bibnamefont{and}
  \bibinfo{author}{\bibfnamefont{O.}~\bibnamefont{Pouliquen}},
  \bibinfo{journal}{\href{https://doi.org/10.1103/PhysRevLett.107.188301}{Physical
  Review Letters}} \textbf{\bibinfo{volume}{107}}, \bibinfo{pages}{188301}
  (\bibinfo{year}{2011}).

\bibitem[{\citenamefont{Dagois-Bohy et~al.}(2015)\citenamefont{Dagois-Bohy,
  Hormozi, Guazzelli, and Pouliquen}}]{DagoisJFM2015}
\bibinfo{author}{\bibfnamefont{S.}~\bibnamefont{Dagois-Bohy}},
  \bibinfo{author}{\bibfnamefont{S.}~\bibnamefont{Hormozi}},
  \bibinfo{author}{\bibfnamefont{{\'E}.}~\bibnamefont{Guazzelli}},
  \bibnamefont{and}
  \bibinfo{author}{\bibfnamefont{O.}~\bibnamefont{Pouliquen}},
  \bibinfo{journal}{\href{https://doi.org/10.1017/jfm.2015.329}{J. Fluid.
  Mech.}} \textbf{\bibinfo{volume}{776}} (\bibinfo{year}{2015}).

\end{thebibliography}
\pagebreak
\clearpage

\onecolumngrid
\begin{center}
\textbf{Puzzling bubble rise speed increase in dense granular suspensions.\\[.2cm]{\textsc{SUPPLEMENTAL MATERIAL}}}\\[.2cm]
{Christopher Madec$^1$, Briva\"el Collin$^1$, J. John Soundar Jerome$^2$, Sylvain Joubaud$^{1,3}$}\\[.1cm]
{\itshape $^1$ Univ Lyon, ENS de Lyon, Univ Claude Bernard, CNRS,\\ Laboratoire de Physique, $46$ all\'{e}e d'Italie, $69364$, Lyon, France}\\
{\itshape $^2$ Universit\'{e} de Lyon, Universit\'{e} Claude Bernard Lyon$1$,\\ Laboratoire de M\'{e}canique des Fluides et d'Acoustique, CNRS, UMR $5509$,\\ Boulevard $11$ Novembre, $69622$ Villeurbanne CEDEX, Lyon, France}\\
{\itshape $^3$ Institut Universitaire de France (IUF)}
\date{\today}\\[1cm]
\end{center}

In this supplemental material, we present movies of the experiments (\S 1), the theoretical derivation of the expression for the dissipated power [Eq. (5)] in the main text (\S 2) and finally, a comparison between our experimental data and results based on Suspension Balance Model as applied to various rheological models (\S 3). Notations are the same as in the above-mentioned paper.

\section*{(1) Supplementary Movies} 
The snapshots presented in Fig.~1(d) correspond to different experimental parameters and are extracted from different movies, which clearly shows the motion of the suspension texture around the rising bubble. The following real-time movies display an experimental front view (width $45$~mm, height $90$~mm) of the experimental Hele-Shaw cell.
\begin{itemize}
\item Rising$\_$bubble$\_$phi$=0$.avi: Air bubble ($d_b=12.5\pm 0.1$~mm) rising in a pure liquid.
\item Rising$\_$bubble$\_$phi$=10$.avi: Air bubble ($d_b=12.5\pm 0.1$~mm)  rising in a suspension (bulk packing fraction $\phi_0=0.1$, particle diameter $d_p=230$~$\mu$m).
\item Rising$\_$bubble$\_$phi$=20$.avi: Air bubble ($d_b=12.5\pm 0.1$~mm) rising in a suspension (bulk packing fraction $\phi_0=0.2$, particle diameter $d_p=230$~$\mu$m).
\item Rising$\_$bubble$\_$phi$=30$.avi: Air bubble ($d_b=12.5\pm 0.1$~mm) rising in a suspension (bulk packing fraction $\phi_0=0.3$, particle diameter $d_p=230$~$\mu$m).
\end{itemize}
\setcounter{equation}{0}
\setcounter{figure}{0}
\setcounter{table}{0}
\setcounter{page}{1}
\renewcommand{\theequation}{S\arabic{equation}}
\renewcommand{\thefigure}{S\arabic{figure}}
\renewcommand{\bibnumfmt}[1]{[S#1]}
\renewcommand{\citenumfont}[1]{S#1}

\section*{(2) Bulk Dissipated Power $P_{\ell}$ in a Fully Developed Pressure-driven Channel Flow of Suspensions.} 
The bulk dissipated power $P_{\ell}$, for $d_{b} \gg h$, is written as
\begin{eqnarray}
\mathcal{P}_{\ell}&=&\displaystyle\int_{\mathrm{cell}} \ \eta_s(y) \left\vert \frac{\partial \mathbf{v}}{\partial y}\right\vert^2 \textrm{d}\tau\\
&=&\displaystyle\int_{d_b/2}^{+\infty}\displaystyle\int_{-\pi}^{\pi}\displaystyle\int_{-h/2}^{h/2} \eta_s(y)\left\vert \frac{\partial \mathbf{v}(r,\theta,y)}{\partial y}\right\vert^2 r\textrm{d}r\textrm{d}\theta\textrm{d}y\,,\label{Power_diss_eq1}
\end{eqnarray}
since ${\mathbf v}(r,\theta,y)=\mathbf{0}$ inside the bubble (see Fig.~4). The integral over the thickness may be rewritten using an integration by parts, since $\mathbf{v}(y=\pm h/2) = \mathbf{0}$:
\begin{equation}
\mathcal{P}_{\ell}=-\displaystyle\int_{d_b/2}^{+\infty}\displaystyle\int_{-\pi}^{\pi}\displaystyle\int_{-h/2}^{h/2} \mathbf{v}(r,\theta,y)\cdot\frac{\partial}{\partial y}\left(\eta_s(y)  \frac{\partial \mathbf{v}(r,\theta,y)}{\partial y}\right)r\textrm{d}r\textrm{d}\theta\textrm{d}y\,,\label{Power_diss_eq2}
\end{equation}
where $\cdot$ denotes the scalar product.
In our case, the Reynolds number based on the bubble speed ($v_b$) and diameter ($d_b$) is smaller than $0.1$ so that we can neglect inertia terms. Since we are also in a Hele-Shaw setup, the leading order viscous term comes only from the second derivatives in the $y$ direction (cell width)~\citep{hele1898flow}. The local velocity field $\mathbf{v}(r,\theta,y)$ is given by
\begin{equation}
\frac{\partial}{\partial y}\left(\eta_s(y)\frac{\partial \mathbf{v}}{\partial y}\right) = \bm{\nabla} P,
\label{eq:NS-HeleShaw}
\end{equation}
where $P$ is the pressure field which depends on $r$ and $\theta$. The dissipated power is, therefore, given by
\begin{eqnarray}
\mathcal{P}_{\ell}&=&-\displaystyle\int_{d_b/2}^{+\infty}\displaystyle\int_{-\pi}^{\pi}\bm{\nabla}P\cdot\left(\displaystyle\int_{-h/2}^{h/2} \mathbf{v}(r,\theta,y)\textrm{d}y\right)r\textrm{d}r\textrm{d}\theta\label{Power_diss_eq3}\\
&=&-h\displaystyle\int_{d_b/2}^{+\infty}\displaystyle\int_{-\pi}^{\pi}\bm{\nabla}P\cdot \mathbf{\tilde{v}}(r,\theta)r\textrm{d}r\textrm{d}\theta\,,\label{Power_diss_eq4}
\end{eqnarray}
where $\mathbf{\tilde{v}}$ is the thickness-averaged velocity field, defined as 
\begin{eqnarray}
\mathbf{\tilde{v}}(r,\theta) = \frac{1}{h}  \displaystyle\int_{-h/2}^{h/2} \mathbf{v}(r,y,\theta) \textrm{d}y. \label{eq:def_v_averaged}
\end{eqnarray}
Since the flow is symmetric about $y=0$, taking $\frac{\partial \mathbf{v}}{\partial y} = 0$ for $y=0$ and integrating Eq.~(\ref{eq:NS-HeleShaw}) leads to
\begin{equation}
\eta_s(y)\frac{\partial \mathbf{v}}{\partial y} = y\bm{\nabla} P\,.
\label{eq:deriv_v}
\end{equation}
A second integration gives the expression of the velocity field as 
\begin{equation}
\mathbf{v}(r,\theta,y)=\bm{\nabla} P(r,\theta)\displaystyle\int_{-h/2}^{y}\frac{y_1}{\eta_s(y_1)}\textrm{d}y_1\,.
\label{eq:v}
\end{equation}
By combining Eq.~(\ref{eq:v}) and Eq.~(\ref{eq:def_v_averaged}), we have
\begin{eqnarray}
\tilde{\mathbf{v}}(r,\theta)=\frac{\bm{\nabla} P(r,\theta)}{h}\displaystyle\int_{-h/2}^{h/2}\displaystyle\int_{-h/2}^{y}\frac{y_1}{\eta_s(y_1)}\textrm{d}y_1\textrm{d}y.
\label{eq:v_averaged}
\end{eqnarray}
Injecting Eq.~(\ref{eq:v_averaged}) into Eq.~(\ref{Power_diss_eq4}), one can get
\begin{eqnarray}
\mathcal{P}_{\ell}=-h^2\frac{\displaystyle\int_{d_b/2}^{+\infty}\displaystyle\int_{-\pi}^{\pi}\vert\mathbf{\tilde{v}}(r,\theta)\vert^2 r\textrm{d}r\textrm{d}\theta}{\displaystyle\int_{-h/2}^{h/2}\displaystyle\int_{-h/2}^{y}\frac{y_1}{\eta_s(y_1)}\textrm{d}y_1\textrm{d}y}.\label{Power_diss_eq5}
\end{eqnarray}
Using the expression of the dipolar field, $\vert\mathbf{\tilde{v}}(r) \vert= v_b/4 (d_b/r)^2$, the numerator can be computed as
\begin{eqnarray}
\displaystyle\int_{d_b/2}^{+\infty}\displaystyle\int_{-\pi}^{\pi}\vert\mathbf{\tilde{v}}(r,\theta)\vert^2 r\textrm{d}r\textrm{d}\theta=\frac{\pi v_b^2 d_b^2}{4},
\end{eqnarray}
while the denominator can be rewritten using an integration by parts
\begin{equation}
\displaystyle\int_{-h/2}^{h/2}\displaystyle\int_{-h/2}^{y}\frac{y_1}{\eta_s(y_1)}\textrm{d}y_1\textrm{d}y=-\int_{-h/2}^{+h/2} \frac{y^2}{\eta_s(y)}\textrm{d}y\,.
\end{equation}
Therefore, the bulk dissipated power is given by
\begin{equation}
\mathcal{P}_{\ell}=\frac{\pi \eta_0 v_b^2 d_b^2 h^2}{4\displaystyle\int_{-h/2}^{+h/2} \frac{y^2}{\hat{\eta}(\phi(y))} \textrm{d}y},
\end{equation}
where we have introduced the suspension viscosity $\eta_s(y) = \eta_0 \hat{\eta}(\phi(y))$ as a function of the suspending liquid viscosity $\eta_0$ and its local volume fraction $\phi(y)$. If the volume fraction $\phi(y) = \phi_0$ is uniform throughout the suspension, one gets
\begin{equation}
\mathcal{P}_{\ell}^0=\frac{3\pi \eta_s(\phi_0) v_b^2 d_b^2}{h}\,.
\end{equation}
Finally, using this expression, the dissipated power is given by
\begin{eqnarray}
\mathcal{P}_{\ell} & = &  \frac{\mathcal{P}_{\ell}^0}{\alpha(\phi_0)} \equiv \dfrac{3\pi d_b^2 \eta_s(\phi_0) v_b^{2}/h}{\alpha(\phi_0)} ,  \\
\alpha(\phi_0) & = & \frac{3}{2} \displaystyle\int_{-1}^{1}\hat{y}^2\frac{\hat{\eta}(\phi_0)}{\hat{\eta}(\phi(\hat{y}))}\mathrm{ d}\hat{y},
\end{eqnarray}
where $\hat{y} = 2y/h$ and $\phi(\hat{y})$ is the local volume fraction distribution. Note that this expression is valid for a circular bubble of diameter $d_b$. A similar calculation for an elliptical bubble of width $D$ and length $L$ [Fig.~1~(c)] leads to the equations~(5) and~(6) of the main text.

\section*{(3) Dissipation-deficit Coefficient $\alpha(\phi_0)$ for Maron-Pierce and also, other well-known Rheological Models.}

\begin{figure}[h]
\includegraphics[width=0.99\textwidth] {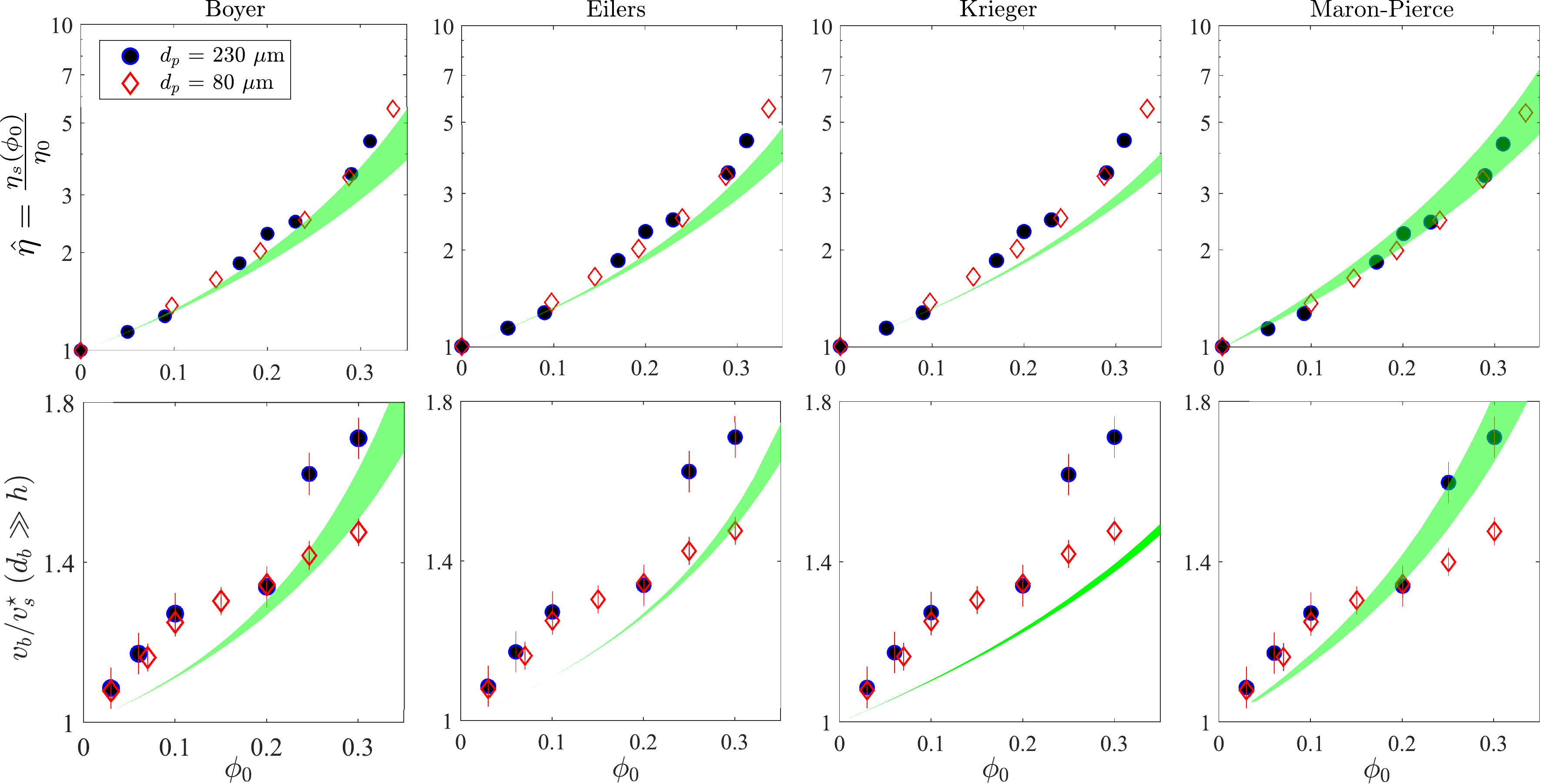}
\caption{\label{fig:AlphaComparison}[top] Same rheological data as in Fig. $1$(c) \& [bottom] same data as in Fig. $3$(c) of the main text showing a comparison between the experimentally measured velocity overshoot $v_b/v_s^\star$ for flat bubbles in a Hele-Shaw cell ($d_b \gg h$) and the computed values of the dissipation-deficit coefficient $\alpha(\phi_0)$ for other commonly used rheological models~\citep{GuazzelliPouliquen2018}, namely, \citet{BoyerPRL2011}, Eilers with $\hat{\eta}(\phi) = \left(1 +  (5\phi/4)/\left(1 -  \phi/\phi_c \right) \right)^{2}$, Krieger with $\hat{\eta}(\phi) = \left(1 -  \phi/\phi_c \right)^{-2\phi_c}$ and Maron-Pierce, which is the one used in the main text.}
\end{figure}
In the recent past, shear-induced particle migration had been successfully modelled by Suspension Balance Model (SBM)~\cite{nott1994pressure} in many configurations ~\citep{morris1999curvilinear, koh1994experimental, yeo2011numerical, snook2016dynamics, sarabian2019}.  In short, for SBM, the normal stress associated with the particle phase arising from collision-dominated agitation of non-Brownian particles in the presence of an external shear rate is the source of the particle migration flux~\citep{nott1994pressure, GuazzelliPouliquen2018}:
\begin{eqnarray}
\dfrac{\partial \phi}{\partial t} = \dfrac{4 \lambda_2 d_p^2}{9 h^2}\dfrac{\partial}{\partial \hat{y}}\left(\left(1 -  \phi \right)^5 \dfrac{\partial}{\partial \hat{y}} \left(\dfrac{\hat{y}}{\mu(\phi)}  \right) \right),
\end{eqnarray}
where $\lambda_2$ is a rheological constant of $\mathcal{O}(1)$ and $\mu(\phi) = \hat{\eta}(\phi) \left(\phi -  \phi_c \right)^2/\phi^2$ is the ratio between the local shear and normal stresses in the suspension. The latter is equivalent to the local ``friction'' coefficient of the suspension~\citep{BoyerPRL2011, DagoisJFM2015, GuazzelliPouliquen2018}. This so-called suspension balance equation allows for a steady-state solution
\begin{eqnarray}
\dfrac{\hat{y}}{\mu(\phi(\hat{y}))} &= &\beta \\
\Rightarrow \dfrac{\phi^2}{\hat{\eta}(\phi) \left(\phi -  \phi_c \right)^2} &= &\dfrac{\beta}{\hat{y}},
\end{eqnarray}
where the singularity at the channel center $\hat{y} = 0$ is avoided by imposing max$(\phi) = \phi_c$ while the constant $\beta$ can be computed by imposing the bulk volume fraction $\phi_0$. Since, for Maron-Pierce rheological formula~\citep{MaronPierce1956} $\hat{\eta}(\phi) = \left(1 -  \phi/\phi_c \right)^{-2}$, it is straight-forward to show that
\begin{eqnarray}
\phi(\hat{y}) = 
	\begin{cases} 
		\phi_c & 0 \leq \vert \hat{y} \vert  \leq \beta, \\
		\phi_c\sqrt{\dfrac{\beta}{\hat{y}}} & \beta \leq \vert \hat{y} \vert  \leq 1, 
   	\end{cases}
\end{eqnarray}
where $\sqrt{\beta} = 1 - \sqrt{1 - \phi_0/\phi_c}$. In fact, a simple closed-form expression for the dissipation deficit coefficient can then be obtained for this particular case, so that
\begin{eqnarray}
\alpha(\phi_0) =  \dfrac{1}{\left(1 - {\phi_0}/{\phi_c}\right)^{2}}\left( 1 - \frac{12}{5}\left(1 - \sqrt{1 - \phi_0/\phi_c}\right) + \frac{3}{2}\left(1 - \sqrt{1 - \phi_0/\phi_c}\right)^2 - \frac{1}{10}\left(1 - \sqrt{1 - \phi_0/\phi_c}\right)^6\right),
\end{eqnarray}
which is precisely the expression for $\alpha(\phi_0)$ that is used in Fig.~$3$(c) of the main text. {Note that $\alpha(\phi_0)$ is larger than $1$ for all bulk volume fraction.} Similar calculations can also be done for different rheological models~\citep{GuazzelliPouliquen2018} for $\hat{\eta}(\phi)$, for example, Eilers with $\hat{\eta}(\phi) = \left(1 +  (5\phi/4)/\left(1 -  \phi/\phi_c \right) \right)^{2}$, Krieger with $\hat{\eta}(\phi) = \left(1 -  \phi/\phi_c \right)^{-2\phi_c}$ and ~\citet{BoyerPRL2011} with $\hat{\eta}(\phi) = 1 +  (5\phi/2)\left(1 -  \phi/\phi_c \right)^{-1} + \{\mu_1 + (\mu_2 - \mu_1)/ [1 + I_0 \phi^2 \left(\phi_c  -\phi \right)^{-2}] \}(\phi/\phi_c)^2\left(1 -\phi/\phi_c \right)^{-2}$ (here, $\mu_1 = 0.32$, $\mu_2 = 0.7$ and $I_0 = 0.005$). Fig. ~\ref{fig:AlphaComparison} shows that irrespective of the rheological model used, the computed $\alpha(\phi_0)$ compares very well the observed velocity-overshoot for various bulk suspension volume fraction in all our experiments.

\end{document}